\documentclass[a4paper,11pt]{article}
\pdfoutput=1 

\usepackage{jcappub}

\usepackage[T1]{fontenc} 
\usepackage{ulem}
\usepackage[dvipsnames]{xcolor}

\newcommand\philin{\phi^{(1)}}

\title{Including relativistic and primordial Non-Gaussianity contributions in cosmological simulations by modifying the initial conditions}

\author[a,b]{Miguel Enríquez,}
\author[b]{Juan Carlos Hidalgo,}
\author[c]{Octavio Valenzuela}

\affiliation[a]{Instituto de Investigación en Ciencias Básicas y Aplicadas, Universidad Autónoma del Estado de Morelos,
\\ 62210, Cuernavaca, Morelos, México}
\affiliation[b]{Instituto de Ciencias Físicas, Universidad Nacional Autónoma de México,
\\ 62210, Cuernavaca, Morelos, México}
\affiliation[c]{Instituto de Astronomía, Universidad Nacional Autónoma de México,
\\ 04510, Coyoacán, Ciudad de México, México}

\emailAdd{miguel.evargas@icf.unam.mx}
\emailAdd{hidalgo@icf.unam.mx}
\emailAdd{octavio@astro.unam.mx}

\abstract{
  We present a method to implement relativistic corrections to the evolution of dark matter structures in Newtonian simulations of a $\Lambda$CDM universe via the initial conditions. We take the nonlinear correspondence between the Lagrangian (Newtonian) evolution of dark matter inhomogeneities and the synchronous-comoving (relativistic) matter density description, and use it to promote the relativistic constraint as the initial condition for numerical simulations of structure formation. In this case, the incorporation of  Primordial non-Gaussianity (PNG) contributions as initial conditions is straightforward. We implement the relativistic, $f_{\rm NL}$ and $g_{\rm NL}$ contributions as initial conditions for the L-PICOLA code, and compute the power spectrum and bispectrum of the evolved matter field. We focus specifically on the case of largest values of non-Gaussianity allowed at 1-$\sigma$ by Planck observations ($f_{\rm NL} = -4.2 $ and $g_{\rm NL} = -7000$). As a checkup, we show consistency with the one-loop perturbative prescription and with a fully relativistic simulation (\texttt{GRAMSES}) on the adequate scales.  
  Our results confirm that both relativistic and PNG features are most prominent at very large scales and for squeezed triangulations. We discuss future prospects to probe these two contributions in the bispectrum of the matter density distribution. \\

}
\begin{document}
\maketitle
\flushbottom
\newpage

\section{Introduction}

 With the aid of technological advances, cosmological observations have arrived to an era of high precision,  where results from Stage IV surveys, such as DESI \cite{desi} or LSST \cite{lsst}, promise to be accurate enough to improve the constraints of the different cosmological models that remain valid. Of particular interest are the constraints to the primordial non-Gaussianity (PNG) parameters, such as the $f_{\rm NL}$ and $g_{\rm NL}$ variables \cite{cmbfnl, pettinari}. Departures from Gaussianity  have been found in the statistics of matter distribution of large scale structure (LSS). The effects have been quantified through numerical simulations on the nonlinear powerspectrum and bispectrum (see e.g. \cite{sefusatti2010,djaques}). In the near future, the statistics of matter density distribution will be determined in great detail, and particularly the high order spectra, such as the bispectrum and trispectrum of matter may contribute to constrain PNG parameters \cite{shirasaki}. 

The expected precision from observations demands equally precise estimations of density distribution in modelling the large scale structure  (LSS) from theoretical models. This requirement has pushed for improvements in numerical simulations of the process of structure formation, both in terms of better resources and more sophisticated algorithms, in order to solve the equations of motion and compute statistical estimators with high enough precision. All of this  with the aim of producing satisfactory observational simulated data to contrast it with the statistics of galaxy surveys \cite{unit}.

Numerical simulations, such as $N$-body, intend to reproduce the full non-linear evolution of the density field, usually assuming Newtonian gravity both to set the initial conditions (trough perturbation theory in a Lagrangian frame) and also for the evolution of the positions and velocities (e.g. the Gadget2 code \cite{gadget}). One $N$-body technique, the particle mesh method (PM), converts the particle distribution to a particle mesh grid with density values. 

In particular, we are interested in the COLA algorithm, (see e.g. \cite{klypin-Holtzmann,lpicolals}), a PM code that solves the Poisson equation through a spectral method using a continuous density field built applying a Cloud-in-Cell (CIC) algorithm to the particle distribution that corrects particle displacements using a 2LPT solution. Besides the speed for producing simulations, another useful aspect of the COLA method is the consistency of the Lagrangian system with a hydrodynamical description at large scales \cite{Oshea,lpicola}. We assume this equivalence to implement relativistic contributions as well as primordial non-Gaussianity through initial conditions for N-body codes. We assume this equivalence to implement relativistic contributions as well as primordial non-Gaussianity.


The main result extracted from numerical simulations is the non-linear evolution of inhomogenities, starting from primordial fluctuations of the early Universe. One of the intrinsically non-linear observables is the matter bispectrum: the Fourier space equivalent of the three-point correlation in the matter distribution. Due to its non-linear character, this observable is expected to show the effects of primordial Non-Gaussianity \cite{sefusatti2010,shirasaki,maartens}, complementing the effects expected in the power spectrum \cite{dalal,mcdonald,brunietaldisentangling,barreira1}. When computed in an evolved matter distribution, the bispectrum also shows the non-linearities of the clustering process, as has been shown in the perturbative regime of the Newtonian \cite{shirata,gallagher} and relativistic theories \cite{umeh,castiblanco,calles,jolicoeur}. 

The non-linearities from the structure formation process induce mode-mixing, that can alter or hide primordial non-linear signatures. In this sense, a cautious implementation for the primordial non-Gaussianity in the initial conditions for numerical simulations, and the correct account of extra contributions is mandatory if one intends to extract the  information pertaining PNG parameters from the evolved matter distribution \cite{dio}.

Several numerical codes have been proposed in order to account for the effects of the full theory of gravity, General Relativity, in the process of structure formation \cite{eloisa,gevolution, gramses}. Yet solving the set of Einstein Field Equations (EFEs) is costly in terms of time and computing resources, and thus ineffective in the determination of parameter values.  

In this paper we propose a method to account for the Primordial non-Gaussianity and the General Relativistic contributions in the matter density distribution as initial conditions for otherwise Newtonian numerical simulations of LSS. Taking heed of the fact that the main differences with the Newtonian description lie in the constraints of the EFEs \cite{bruni,brunihidalgowands,bertaccaetal}, previous works have presented the relativistic contributions to the polyspectra of the evolved dark matter at the perturbative level \cite{gressel-bruni,rebecca}, and interpreted the relativistic constraint as the initial displacement of dark matter particles in Lagrangian coordinates \cite{christopherson}. Such method represents a low-cost implementation of the dominant relativistic effects in the matter distribution, together with the signature of PNG. These findings lead us to interpret the numerical evolution of Lagrangian displacements, like the L-PICOLA code, as an algorithm to evolve the matter field consistently within GR. We thus show how to set relativistic and non-Gaussian initial conditions in the L-PICOLA code.  In practice, to implement the initial conditions we employ the 2LPTIC code \cite{2lptic} and modify the extension originally produced to incorporate PNG \cite{scomanera}. The ultimate goal of this paper is to provide a quantitative difference in the evolved matter density between the Gaussian and the non-Gaussian field with relativistic contributions, and discuss its detectability in future galaxy surveys.

The paper is organized as follows: in Section~\ref{sec:pertIC} we review the perturbative formalism to account for the different contributions to the density contrast. In Section~\ref{secini} we describe the implementation from the non-linear density contrast to the gravitational potential kernel to be included in the numerical simulation codes. We present in this section (Section~\ref{sec:oneloop}) the one-loop power spectrum and the tree-level bispectrum from perturbation theory in order to show consistency with the numerical results. In Section~\ref{results} we present the power spectrum and bispectrum computed with the Pylians code \cite{pylians} and a comparison between simulations with  relativistic, non-Gaussian and Gaussian initial conditions. Finally, in Section~\ref{discussion} we discuss the results and the prospects of future work.

\section{Perturbative Initial Conditions}
\label{sec:pertIC}
\subsection{Evolution equations}
The synchronous-comoving gauge is often chosen \cite{sinc} for evolution equations for the density contrast to have a suitable Lagrangian frame in GR and for defining local Lagrangian galaxy bias up to second order. It defines the line element as
\begin{equation}
    ds^2=a^2(\eta)[-(1+2\phi)d\eta^2+2\omega_{,i}d\eta dx^i+\gamma_{ij}dx^idx^j] \ \ \ ,
\end{equation}
where $a$ is the scale factor, $\eta$ is the conformal time. $\phi$ and $\omega$ are scalar metric perturbations and $\gamma_{ij}$ is the spatial metric, with latin indices accounting for the three spatial coordinates. Working in the synchronous-comoving, we set $\phi=\omega_{,i}=0$.
The matter considered, is a pressureless fluid as described by comoving observers. This defines the four-velocity as $u_\mu=(-a,0,0,0)$ (with greek indices running from 0 to 3 and representing spacetime coordinates). So the deformation tensor is \cite{ellis},
\begin{equation}
    \vartheta^\mu_{~\nu} \equiv a u^\mu_{~;\nu}-\mathcal{H} \delta^{\mu}_{~\nu} \ \ \ .
\end{equation}
This tensor presents only spatial components proportional to the extrinsic curvature $K_j^i$ of the conformal spatial metric $\gamma_{ij}$
\begin{equation}
    \vartheta^i_ j=-K^i_j \ \ \ ,
\end{equation}
where the extrinsic curvature is given by \begin{equation}
    K^i_j \equiv -\frac{1}{2}\gamma^{ik}\gamma'_{kj}\ \ \ ,
\end{equation}
(where a prime stands for $' \equiv \partial / \partial \eta$).

The above is the basis for the covariant fluid approach to perturbation theory \cite{ellisbruni89,bruni92}. The density field $\rho$ can be split in a density background  $\bar{\rho}(\eta)$ and a fluctuation $\delta\rho(\bold{x},\eta)$ is 
\begin{equation}
    \rho(\bold{x},\eta)=\bar{\rho}(\eta)+\delta\rho(\bold{x},\eta)=\bar{\rho}(\eta)(1+\delta(\bold{x},\eta)) \ \ .
\end{equation}
The continuity equation for the evolution of the density contrast $\delta(\mathbf{x},\eta)$ is
\begin{equation}
\label{cont}
    \delta'+(1+\delta)\vartheta =0 \ \ \ ,
\end{equation}
where $\vartheta=\vartheta_{\alpha}^{~\alpha}$ is the trace of $\vartheta_\nu^{~\mu}$.
The evolution of $\vartheta$ is described by the Raychaudhuri equation which is
\begin{equation}
\label{ray}           
    \vartheta'+\mathcal{H}\vartheta+\vartheta_{~j}^i\vartheta_{~i}^j+4\pi Ga^2\bar{\rho}\delta=0\ \ \ .
\end{equation}
The Raychaudhuri equation is analog to Euler's equation in the GR regime which describes the motion of particles \cite{ellis}.  
Equations \eqref{cont}~and~\eqref{ray} describe the evolution of a dust component in a cosmological de-Sitter or $\Lambda$CDM background. Note that these two equations find a non-linear equivalence with the Newtonian continuity and Euler equations in the Lagrangian frame, when the following equivalences are drawn (see e.g. \cite{ellisbruni}): 
\begin{equation*}
    Newtonian\ Lagrangian\ \longleftrightarrow\ Relativistic\ comoving
\end{equation*}
\begin{equation*}
    \frac{d}{dt} \longleftrightarrow \frac{\partial}{\partial \eta}
\end{equation*}
\begin{equation*}
    \partial^i v_j \longleftrightarrow \vartheta^i_ j 
\end{equation*}
\begin{equation*}
    \delta_N \longleftrightarrow \delta
\end{equation*}

After the consideration of a geometrical equivalence between the deformation tensor and the Ricci curvature, the 00-component of the Einstein field equation, the energy constraint, can be written as \cite{ellisbruni89}
\begin{equation}
\label{enercon}
    \vartheta^2-\vartheta_{~j}^i\vartheta_{~i}^j+4\mathcal{H}\vartheta+^3R=16\pi Ga^2\bar{\rho}\delta \ \ \ ,
\end{equation}

A similar equation can be derived from the Newtonian conservation of energy equation with $E_N~=~1/2 v^2 - \phi_N$.
In that case one can show that, at first order, the pertrurbative part of the Newtonian (conserved) energy is equal to the spatial curvature  ${}^{3}R^{(1)} = -4 \nabla^2 \delta E_N^{(1)}$.

At non-linear order the correspondence with the Newtonian energy conservation is broken but the Ricci three-curvature is still time-independent and can be expanded in terms of the (non-linear) metric potentials encoded in $\gamma_{ij}$. 

At large scales, the differences between the Newtonian and relativistic descriptions of the non-linear inhomogeneities lie within the constraint equation, and are dominated by the spatial curvature term.
This is justified in the following through a gradient expansion. In particular, we show that the curvature terms are dominant at early times in the relativistic constraint \eqref{enercon}. As we shall see, such contributions also include the primordial non-Gaussianity if present.

\subsection{Perturbative and gradient expansion of the relativistic contribution}
\label{pert}

Restoring to a perturbative expansion, can be written the scalar quantities as for example,      \begin{equation}
\label{expan}
        \delta = \delta^{(1)}+\frac{1}{2}\delta^{(2)}+\frac{1}{6}\delta^{(3)}+ \ldots
\end{equation}

\noindent where the superscript in parenthesis indicates the perturbative expansion order.

 For the linear order, the solution for the density contrast is
 \begin{equation}
     \delta^{(1)}=\frac{D_{+}(\eta)}{10\mathcal{H}_{IN}D_{+IN}}(-4\nabla^2\zeta^{(1)})  \ ,
 \end{equation}
 where the growth factor in Eintein-de Sitter Universe is \cite{bernardeau}
 \begin{equation}
     D_{+}=\frac{D_{+IN}\mathcal{H}_{IN}^2}{\mathcal{H}^2} \ ,
 \end{equation}

\noindent and where the subindex $IN$ represents values at an arbitray initial time in the matter-dominated universe.

The curvature perturbation $\zeta$ is one of two scalar degrees of freedom that are encoded in the spatial metric. Considering the synchronous-comoving gauge, the expansion for $\gamma_{ij}$ is
\begin{align} 
\notag
    \gamma_{ij} =& \exp[2\zeta]\delta_{ij} +
\left(\partial_i\partial_j  - \frac{1}{3} \nabla^2\right) \chi \\ =& \left[1 + \zeta^{(1)} +  2\zeta^{(1)2} +\zeta^{(2)} + (\partial_i\partial_j  - \frac{1}{3} \nabla^2)\left(\chi^{(2)} + \frac{1}{2}\chi^{(2)}\right) \right] + \ldots 
\end{align}

\noindent where $\chi$ encodes the traceless perturbation of the spatial metric. 
The large scales contributions are dominated by the conformal perturbation $\zeta$, which corresponds to the curvature perturbation in the uniform density gauge \cite{bruni}, then it can be written as
\begin{equation}
    g_{ij}=a^2\gamma_{ij}=a^2e^{2\zeta}\bar{\gamma}_{ij} \ \ \ .
\end{equation}

The above can be used to set initial conditions in the early Universe after inflation. In such scenario, the curvature perturbation $\zeta$ is nearly scale-invariant and remains constant. Then, in order to capture the contribution of relativistic terms, we perform a gradient expansion (long-wavelength approximation), where the spatial gradients are small compared to time derivatives. We thus note that the following perturbed quantities are of second order:
\begin{equation}
\label{grad}
    \delta\sim \vartheta\sim {}^3R \sim \nabla^2 \ \ \ .
\end{equation}

If we stick to the large-scales in a gradient expansion in powers of $\nabla/\mathcal{H}$, which is the regime where the relativistic corrections are expected to be significant, then it can be shown that the dominant contribution to the three-curvature comes from the conformal metric potential $\zeta$.
Consequently we can safely adopt the approximation $\bar{\gamma}_{ij}\simeq\delta_{ij}$. Thus the Ricci scalar takes the form \cite{brunihidalgowands,gressel-bruni}
\begin{equation}
    {}^3R=-4\nabla^2\zeta+\sum_{m=0}^{\infty}\frac{(-2)^{m+1}}{(m+1)!}[(m+a)(\nabla\zeta)^2-4\zeta\nabla^2\zeta]\zeta^m
    \label{curvature}
\end{equation}
The curvature perturbation expanded in terms of a Gaussian random field $\zeta^{(1)}$ is (see e.g. \cite{wands2014} for the gauge-invariant expression of non-Gaussianity parameters):
\begin{equation}
\label{zeta}
    \zeta=\zeta^{(1)}+\frac{3}{5}f_{\rm NL}\zeta^{(1)2}+\frac{9}{25}g_{\rm NL}\zeta^{(1)3} \ .
\end{equation}
With the previous expansion, one can obtain solutions for the density contrast at higher orders. For $m=1$ in Eq.~\eqref{curvature} the third order corrections are as follows
\begin{equation}
    {}^3R=-4\nabla^2\zeta+(-2)[(\nabla\zeta)^2-4\zeta\nabla^2\zeta]+2[2(\nabla\zeta)^2-4\zeta\nabla^2\zeta]\zeta \ \ \ .
\end{equation}
The Ricci scalar at large scales (at second order in a gradient expansion) and considering scalars only, is given in terms of the Gaussian field as
\begin{equation}
\begin{aligned}
{\ }^{3}R\simeq -4\nabla^2\zeta^{(1)}+& \left( \nabla\zeta^{(1)} \right)^2\left[ -2-\frac{24}{5} f_{\rm NL} \right] + \zeta^{(1)}\nabla^2 \zeta^{(1)} \left[ -\frac{24}{5}f_{\rm NL}+8\right]\\  + &\zeta^{(1)}\left( \nabla\zeta^{(1)} \right)^2\left[ -\frac{216}{25}g_{\rm NL}+\frac{24}{5}f_{\rm NL}+4\right]\\  + &\zeta^{(1)2}\nabla^2 \zeta^{(1)}\left[ -\frac{108}{25}g_{\rm NL}+\frac{72}{5}f_{\rm NL}-8\right]+O(\zeta^{(1)4}) \ \ \ .
\end{aligned}
\label{ricci}
\end{equation}

This expression sums up the justification for the present work: The three-curvature ${}^3R$ encodes the differences, at the level of constraints, between the Newtonian and the Relativistic formalisms of structure formation, since the evolution equations \ref{cont} and \ref{ray} for a cold dark matter component are identical at non-linear level between these two formalisms (at large scales, where the gradient expansion remains valid). This means that Relativistic initial conditions included in ${}^3R$ can be evolved employing a Newtonian hydrodynamical code, or its equivalents, and the results remain consistent with GR. Moreover,  we notice that even in the case that initial conditions limited to relatively large scales, i.e., at lowest order in the gradient expansion $\nabla/\mathcal{H}$, we can still express the relativistic contributions in terms of $\zeta$, and also allow for a non-linear primordial $\zeta$, that is, to include primordial non-Gaussianities as those expressed in Eq.~\eqref{zeta}.

These considerations argue that the formalism presented can be implemented in hydrodynamic codes (e.g.~\cite{ramses} or \cite{arepo}) or codes which show equivalence with the hydrodynamical description at large scales, such as L-PICOLA \cite{lpicola}, which we discuss in more detail in the following Sec.~\ref{sec:sims}. This allows for the evolution of initial conditions which include relativistic constraints through Newtonian equations\footnote{expressions for the Lagrangian displacement with relativistic input have been discussed in previous works, \cite{christopherson,ismael}. These can be regarded as strategies to modify the initial conditions generating code to support directly the grid-based setup.}.

To end this section, we present explicitly the matter density field up to third order in perturbation theory. This represents the homogeneous part of the solution to the constraint~\eqref{enercon}  (valid at second order in a gradient expansion) \cite{gressel-bruni,rebecca}. 
\begin{equation}
    \frac{1}{2}\delta^{(2)}=\frac{D_{+}(\eta)}{10\mathcal{H}^2 D_{+IN}}\frac{24}{5}\left[ -(\nabla\zeta^{(1)})^2\left(\frac{5}{12}+f_{\rm NL}\right)+\zeta^{(1)}\nabla^2\zeta^{(1)}\left(\frac{5}{3}-f_{\rm NL}\right)\right] \ \ \ ,
\label{del2}
\end{equation}    
\begin{equation}
\begin{aligned}
    \frac{1}{6}\delta^{(3)}=\frac{D_{+}(\eta)}{10\mathcal{H}^2_{IN}D_{+IN}}
    \frac{108}{25}&\biggl[   2\zeta^{(1)}(\nabla \zeta^{(1)})^2 \left( -g_{\rm NL} + \frac{5}{9}f_{\rm NL}+\frac{25}{54} \right) \\      & +\zeta^{(1)2}\nabla^2\zeta^{(1)} \left(-g_{\rm NL}+\frac{10}{3}f_{\rm NL}-\frac{50}{27} \right) \biggr] \ \ .
\end{aligned}
\label{del3}
\end{equation}

In the following we implement these solutions as initial conditions for a numerical simulation in order to compute the the polyspectra of the evolved density field for cases of interest. We also employ these last two expressions to complement the usual (Newtonian) one-loop contributions to the power spectrum as detailed in Ref.~\cite{rebecca} and implement them as discussed in Sec.~\ref{sec:oneloop}.
\section{Initial Condition implementation and Numerical Evolution}
\label{secini}
    In order to introduce a set of initial values of the density field suitable for L-PICOLA, we modify the 2LPTic code \cite{2lptic}, and more specifically
    the extension in \cite{scomanera}, originally produced to include PNG initial conditions. Our modification incorporates the relativistic solutions presented earlier. 
    Specifically, we implement a modification to the kernel of the gravitational potential which meets Poisson's equation at all perturbative orders 
\begin{equation}
    \nabla^{2}\phi=\frac{3}{2}\mathcal{H}^{2}\Omega_m\delta \ \ \ ,
    \label{poisson}
\end{equation}
where $\delta$ is given by the expression \ref{expan}.
The Newtonian potential coincides with the comoving curvature perturbations at first-order, so that at linear level we write
\begin{equation}
    \philin=\frac{3}{5}Rc \approx - \frac{5}{3} \zeta^{(1)}\, .
    \label{philin}
\end{equation}

\noindent in an equivalence valid for the large scales and early times, where our initial conditions are set.

Since the initial conditions code assumes the Poisson equation as valid at all orders, we introduce to that equality the expressions for the Fourier-space equivalent of the density contrast at second and third order in Eqs.~\ref{del2} and \ref{del3}. By rewriting these contributions in terms of the linear potential $\philin$ we can express the non-linear initial potential as a sum of kernels up to third order as
\begin{equation}
    \phi_{\rm ini}=\philin + \frac{1}{2}\phi^{(2)}+ \frac{1}{6}\phi^{(3)} \ \ \ ,
\end{equation}

\noindent with
\begin{equation}
\label{phi2}
    \phi^{(2)} =-\frac{72}{625}\left[(\nabla\philin)^2(\frac{5}{12}+f_{\rm NL})+\philin\nabla^{2}\philin (\frac{5}{3}-f_{\rm NL})\right]
\end{equation}
\begin{equation}
\label{phi3}
\phi^{(3)} = -\frac{972}{15625}\left[ 2\philin (\nabla \philin)^2   \left( g_{\rm NL} - \frac{5}{9}f_{\rm NL}-\frac{25}{54} \right) +\philin{}^2\nabla^2\philin \left(g_{\rm NL}-\frac{10}{3}f_{\rm NL}+\frac{50}{27} \right)\right]
\end{equation}

It is through these expressions that we modify the 2LPTic code using its own aliases for all of the terms contributing to the matter initial conditions. 

\subsection{Cosmological simulations}
\label{sec:sims}

L-PICOLA is a particle mesh (PM) code for gravitational evolution in Lagrangian coordinates, calculating the large-scale dynamics accurately using the 2LPT approximation, while  the N-body code solves for the small scales.
This results in much faster integrations, only for large scales \cite{winther}. The advantages of the COLA method have been exploited to produce fast galaxy mock catalogues both in the standard $\Lambda$CDM model \cite{lpicola,ref8} as well as in models of modified gravity \cite{fastmocks,MGPICOLA}, dynamical dark energy \cite{devi2020}, massive neutrinos \cite{colaneutrinos}, and others.

As with many $N$-body codes, the particles are placed in a three-dimensional uniform grid, and are displaced to their initial position and velocity by the 2LPTic algorithm, thus establishing the initial conditions for the numerical evolution  \cite{2lptic}. Primordial non-Gaussianity has been implemented through an extension of the 2LPTic code which originally introduced second-order terms to the Newtonian potential \cite{scomanera}. We have modified this non-Gaussian version of the code to account for the relativistic contributions of Eqs.~\eqref{phi2}~and~\eqref{phi3}, including the third order PNG terms, which for the first time feature the $g_{\rm NL}$ parameter. We thus argue that the combination of a modified 2LPTic plus the (Newtonian) L-PICOLA evolution represent a suitable frame and fast method to introduce the dominant relativistic contributions from the energy constraint in the synchronous-comoving frame.

For each simulation, we consider $1024^3$ DM particles in a cubic box with side length $L=2048~\mathrm{Mpc}/h$  in order to cover a large enough volume and probe the scales relevant for relativistic effects. These are evolved from redshift $z=49$, In agreement we previous studies \cite{dav97,sefusatti2010,ban18} with similar boxes and resolutions we started the simulation at z= 49. This initial time guarantees that matter fluctuations at the smallest scales sampled ($\approx 2$ Mpc), lie still within the linear regime \cite{gao05}. Each simulation stops at $z=1$ after 150 timesteps, well above the number of steps recommended for L-PICOLA \cite{lpicola}. The cosmological parameters used were taken from Planck data \cite{planckparams,planckng} including the values allowed at 1-$\sigma$ for the  non-Gaussianity parameters $f_{\rm NL}$ and $g_{\rm NL}$. Explicit numbers are displayed in Table \ref{tabb}. We ran 10 simulations with and without the PNG input (both with the relativistic initial condition), and another 10 simulations with Gaussian initial conditions and no relativistic input, which in our case represent simulations with purely linear initial conditions. The analysis of the simulations output is presented in the next section.

\begin{table}[h!]
    \caption{Cosmological parameters}
    \centering
    \begin{tabular}{|c|c|}
\hline
 $\Omega_\Lambda$ & 0.69 \\ 
 \hline
 $\Omega_M$ & 0.31  \\ 
 \hline
 $\Omega_b$ & 0.048 \\
 \hline
 $H_0$ & 0.69  \\
 \hline
 $f_{\rm NL}$ & -4.2 \\
  \hline
 $g_{\rm NL}$ & -7e3 \\
  \hline
 $\sigma_8$ & 0.8 \\
 \hline
 $n_s$ & 0.96  \\ [1ex] 
 \hline
    \end{tabular}

    \label{tabb}
\end{table}
\subsection{ The perturbative one-loop power spectrum and bispectrum}
\label{sec:oneloop}

As a check-up for the consistency of our simulations at the largest scales, we compare the output power spectrum with the perturbative matter power spectrum at one-loop. Both methods should present the same signal at large scales, while differences are expected at the quasi-linear scales and smaller. The perturbative spectrum is given by
\begin{equation}
P(k,\eta)=P_L+2P^{(1,3)}(k,\eta)+P^{(2,2)}(k,\eta) \ \ \ ,    
\end{equation}
where $P_L$ corresponds to the linear contribution and $P^{(1,3)},P^{(2,2)}$ are the one-loop corrections with contributions of the perturbative second and third order respectively. These can be written as
\begin{equation}
P^{(2,2)}(k,\eta)=2\int \frac{d^3q}{(2\pi)^3}P_L (q,\eta)P_L(|\bold{k}-\bold{q}|,\eta)[F^{(2)}(\bold{q},\bold{k-q},\eta)]^2 \ \ ,
\end{equation}
\begin{equation}
    P^{(1,3)}(k,\eta)=3F^{(1)}(\bold{k})P_L(k,\eta)\int \frac{d^3q}{(2\pi)^3}P_L (q,\eta)F^{(3)}(\bold{k,q,-q},\eta) \ \ .
\end{equation}
where $F^{(2)}$ and $F^{(3)}$ are kernels with Newtonian and relativistic contributions at second and third order in Fourier space in the Eulerian frame \cite{gressel-bruni,rebecca}.

Additionally, the tree-level bispectrum is defined as
\begin{equation}
    B(k_1,k_2,k_3)=2P_L(k_1,\eta)P_L(k_2,\eta)F^ {(2)}(\bold{k_1},\bold{k_2})+(2\ \text{cyclic.}) \ \ ,
    \label{bisbis}
\end{equation}
where the kernel $F^{(2)}$ is that of the one-loop power spectrum.

\begin{figure}[t!]
\centering
\includegraphics[scale=0.7]{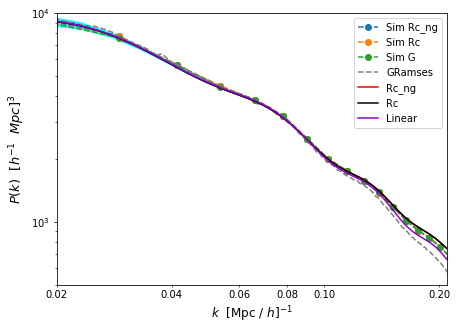}
\caption{Power spectrum from one-loop (lines) and simulations (dots) for relativistic ($Rc$) contributions, for relativistic plus non-Gaussian contributions ($Rc\_{ng}$), and Gaussian (i.e. linear; $G$) initial conditions. The gray line shows a \texttt{GRamses} simulation from \cite{gramses}. The turquoise shaded area represents the variance of a survey with an observable area of $A=14\ 000$deg$^2$ (DESI-like).}
\label{pk}
\end{figure}

\section{Results}
\label{results}

We computed the power spectrum and bispectrum from each of the simulations described previously, with the aid of the script \textsc{Pylians} (Python libraries for the analysis of numerical simulations) \cite{pylians}.     
In order to compare the three studied cases, namely the non-Gaussian relativistic, the Gaussian relativistic, and the Gaussian Newtonian case. For all cases, with used the same seed in such a way residuals are due to the intrinsic differences between the models.

\noindent In Figure \ref{pk} we present the power spectrum at $z = 1$ for the  L-PICOLA output considering the kernel of Section~\ref{secini} as initial conditions, and the corresponding one-loop power spectrum from cosmological perturbation theory. The 3 cases presented are, $Rc$: relativistic contributions with $f_{\rm NL}=0$  $g_{\rm NL}=0$, $Rc \_ng$: relativistic contributions with $f_{\rm NL}=-4.2$ $g_{\rm NL}=-7000$,  and $G$: The reference Gaussian initial conditions. The plotted scales are limited by the fundamental frequency (The box size ) $f_{\rm fun}=2\pi /Lenght$, which in our specific case is $f_{\rm fun}=0.002\ h/Mpc$.

The shaded area in this figure represents the variance expected in an DESI-like survey \cite{desi} for an observable area of $A=14\ 000$deg$^2$ (no particular geometry is considered). We take the inverse of the squared Fisher matrix of eq. \eqref{fish} to compute the variance given a certain volume  \cite{alkis}; that is,
\begin{equation}
    F_{i,j}=\frac{V_s}{4\pi^2}\int_{-1}^1d\mu \int_{k_{min}}^{k_{max}}dk\frac{\partial P(k,\mu)^S}{\partial p_i}\left( P(k,\mu)+\frac{1}{n} \right)^{-2}\frac{\partial P(k,\mu)^S}{\partial p_j} \ \ \ .
    \label{fish}
\end{equation}

\begin{figure}[b!]
\centering
\includegraphics[scale=0.7]{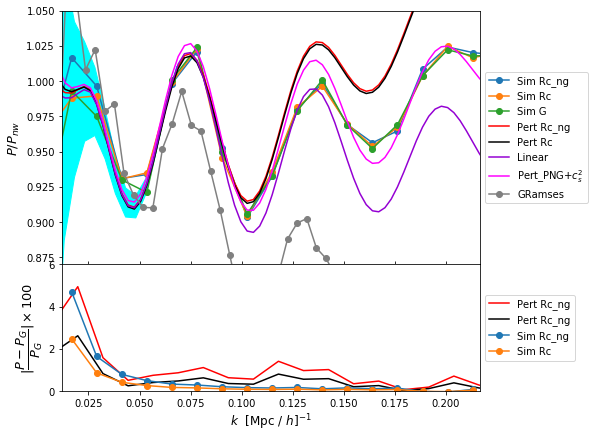}
\caption{Top panel: Power spectrum from one-loop (lines) and simulations (dots) normalized to the no-wiggle power spectrum. Initial conditions are set to Gaussian (G) or including relativistic contributions with ($Rc\_ng$) and without ($Rc$) primordial non-Gaussianity. The gray line is a GRamses simulation from \cite{gramses}. The turquoise shade is the same as in Fig.~\ref{pk}. The pink line represents a one-loop perturbative powerspectrum corrected through a counter-term addition. Bottom panel: the percentage difference between relativistic and Gaussian cases.}
\label{pkfrac}
\end{figure}

\begin{figure}[h!]
\centering
\includegraphics[scale=0.7]{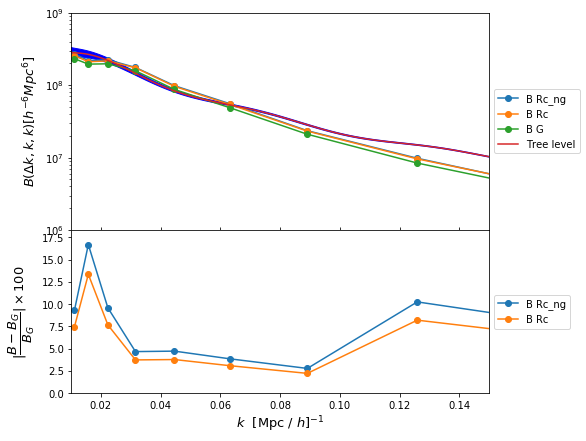}
\caption{\textbf{Bispectrum sensitivity to GR and PNG effects.} Top panel: Bispectrum computed at tree-level in relativistic perturbations (red line, Eq.~\eqref{bisbis}), and from numerical simulations (averaged over 10 realizations) of Gaussian (green dots), relativistic ICs (Orange dots) and the relativistic non-Gaussian (blue dots) cases, at $z=1$, the blue shadow is the variance from a DESI-like survey. Bottom panel: the percentage difference of the bispectrum for relativistic initial conditions simulations with respect to the Gaussian case. ; $Rc_{ng}$ is for relativistic corrections and non-Gaussianity, $Rc$ for just relativistic corrections. }
\label{bb}
\end{figure}

We have included in Figures~\ref{pkfrac}~and~\ref{bb} the data from the full relativistic code \texttt{GRamses}~\cite{gramses}. The suppression observed for the amplitude at low scales is due to the low resolution of that simulation, but we find a good agreement for large scales with our power spectrum (within the errors, and up to $k~0.07$), where the dominant effect takes place. 

In Figure \ref{pkfrac} we compare the perturbative power spectrum at one-loop, as well as our simulations against the no-wiggle power spectrum. In the lower panel where we present the difference with respect to the Gaussian simulations (green dots). On large scales we find an increment up to 4\% (2\%) in the amplitude of the power spectrum in relativistic simulations with (without) non-Gaussianity. At smaller scales, i.e. $k \gtrsim 0.1~h~\mathrm{Mpc}^{-1}$,  find that the discrepancy with the Gaussian simulation vanishes, and recover the well-known divergence of the one-loop perturbative estimation (shown, for example, in Refs.~\cite{carlson09,vlah15}).

In order to ease the small-scale differences, we have included the counterterm $c_s^2$ of effective field theory (see e.g.~ \cite{counterfonseca}) to match the perturbative power spectrum with the numerical (Gaussian) simulations and make it consistent, specially for the small scales. We adopt the counterterm contribution to the power spectrum at lowest-order from EFT (Effective field theory), defined as:
\begin{equation}
    P_{ctr,1loop}\equiv -2k^2c_s^2P_{11}.
\end{equation}

As is usual, the counterterm coefficient was calculated after subtracting the one-loop power spectrum from the mean power spectrum from the simulations \cite{counterbaldauf}. The resulting value shown in Figure \ref{pkfrac} is $c_s^2=12.4$. With the one-loop power spectrum corrected in this way we find that the differences with simulations are below 2 percent throughout the range of scales of Figure \ref{pkfrac}, and that simulations reproduce correctly the one-loop effect of relativistic and non-Gaussianity contributions at large scales.

\noindent With the power spectrum in good agreement with the theoretical prediction and fully relativistic simulations, we explore the effects of our model in the bispectrum. For local non-Gaussianities and at leading order in relativistic effects, we find that the amplitude of the bispectrum is most prominent in the \textit{squeezed} configuration \cite{mata}. In Figure~\ref{bb} we present the bispectrum as a function of scale, for triangles with sides $k_1=k_2=k$ and $k_3=0.013k$. Also, the shaded area around the tree-level bispectrum in this figure represents the variance expected in an DESI-like survey \cite{desi} for an observable area of $A=14\ 000$deg$^2$ (no particular geometry is considered) \cite{oliverbis}.
The lower panel of that figure shows the relative difference with respect to the Gaussian simulations (green dots). We observe that the difference increases with the size of the triangle. The percentage deviations from the Gaussian amplitude are of order 10\% (8\%), for the relativistic initial conditions with (without) non-Gaussianity. Note that at small scales, the PNG contribution vanishes, with a feature at small scales also reported in the literature (see e.g. \cite{sefusatti2010}). Once again, for  $k \gtrsim 0.08~h~\mathrm{Mpc}^{-1}$ the perturbative analysis is divergent from the results of simulations. 
    
\begin{figure}[h!]
\centering
\includegraphics[scale=0.7]{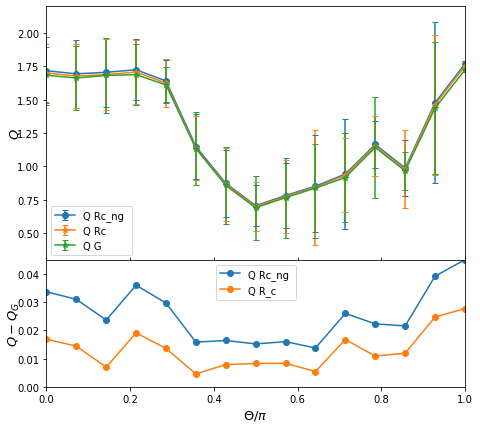}
\caption{\textbf{Reduced bispectrum sensitivity to GR and PNG effects.} Top panel:  reduced bispectrum (Eq.~\eqref{reduced:bis}) for the mean of the 10 realizations of Gaussian simulations ($Q_G$) and  non-Gaussian simulations ($Q_R$ and $Q_N$)  at $z=1$ with $k_1=k_2=0.01$. Bottom Panel: Difference of the reduced bispectrum for non-Gaussian and relativistic IC  simulations with respect to the Gaussian ones.}
\label{q}
\end{figure}

\noindent In Figure~\ref{q} we show  the reduced bispectrum defined as:
\begin{equation}
\label{reduced:bis}
    Q=\frac{B(k_1,k_2,k_3)}{P(k_1)P(k_2)+P(k_2)P(k_3)+P(k_3)P(k_1)} \ \ \ \ ,
\end{equation}
\noindent The bottom panel of that figure shows the difference of the studied cases with respect to the Gaussian ICs. We can see that the difference is most prominent at squeezed configurations (low values of the aperture angle $\theta$), which is the reason behind the choice of triangle shape in the plots of Figure~\ref{bb}.

 \section{Discussion}
 \label{discussion}

 In this paper we have produced L-PICOLA simulations with an input from purely relativistic contributions as well as from primordial non-Gaussianity, introduced through a modification to the initial conditions provided by the 2LPTic code. The input is given in terms of the gravitational potential kernel, which for our work takes corrections at 1-loop including the relativistic terms and primordial non-Gaussianities from $f_{\rm NL}$ and $g_{\rm NL}$ on the local configuration. Using the limit values of the PNG parameters, as imposed by Planck satellite observations, we find higher percentage deviation from the Gaussian case in the bispectrum (as shown in figures~\ref{bb}~and~\ref{q}) than for the power spectrum (Figure~\ref{pkfrac}). Similar differences are obtained when considering exclusively relativistic contributions without primordial non-Gaussianity. The most significant deviations from the Gaussian case were detected at larges scales ($k<0.05$), and for the squeezed triangle configurations, as expected.

The consistency of the numerical simulations with the 1-loop analytic corrections, and with fully relativistic simulations, in the matter power spectrum, allow us to extend the analysis of a relativistic and non-Gaussian signal to smaller scales. The presented technique can, of course, be used to include non-Gaussianities in other configurations (e.g. equilateral), via the 2LPTic prescriptions. We thus have at hand a practical tool to incorporate both the relativistic contributions as well as the correct (gauge-invariant) input from Primordial non-Gaussianity in the initial conditions of codes like L-PICOLA and GADGET-2, as well as other Nbody codes (e.g.~\cite{ramses,fastpm}), and even hydrodynamical codes (e.g.~\cite{arepo}), in order to capture the effects in the large scale structure observables. Our modification will be made public in a GitHub repository in \textit{ https://github.com/miguelevargas/lpicola\_mod} and is available upon request.

 
  While the above results remain within the variance error of the present galaxy surveys, a more refined error analysis can be employed in order to have a better control of errors for the power spectrum and bispectrum. For example, one can use simulations with inverted phases and look at the relation of the power spectrum and its statistical properties with the covariance matrix \cite{klypin18}. 
 There is a variety of applications for our numerical method. One can compute, after several realizations, the covariance matrix of the relativistic and non-Gaussian matter spectrum, and observe the scale and amplitude of mode-mixing \cite{chan,ebossbis,percivalcov}. On the other hand, through a galaxy occupation method (such as HOD \cite{hod}), one can probe the galaxy bias parameters which modulate the contribution of the relativistic and PNG terms in the powers spectrum (see e.g. \cite{mcdonald,barreira1,rebecca2}) and forecast constraints to PNG parameters from the resulting polyspectra (see e.g. \cite{ashelyross}). All these tasks and improvements will be studied in future work.  

\acknowledgments
 We are grateful to Cristian Barrera-Hinojosa for providing data from runs of the GRAMSES code. We also thank Josué de Santiago for assistance in the implementation of the perturbative methods. 
 We acknowledge support through computational and human resources provided by the LAMOD-UNAM project through the clusters Atocatl and Tochtli. LAMOD is a collaborative effort between the IA, ICN, and IQ institutes at UNAM and DGAPA UNAM grants PAPIIT IG101620 and IG101222.
This work is sponsored by CONACyT grant CB-2016-282569 and by Program UNAM-PAPIIT Grant IN107521 "Sector Oscuro y Agujeros Negros Primordiales".

\bibliographystyle{ieeetr}


\end{document}